\documentstyle[sprocl,epsfig]{article}

\def\be{\begin{equation}}
\def\ee{\end{equation}}
\def\bea{\begin{eqnarray}}
\def\eea{\end{eqnarray}}
\begin{document}

\title{  EXTERNAL COULOMB AND ANGULAR 
         MOMENTUM INFLUENCE ON ISOTOPE COMPOSITION OF NUCLEAR 
         FRAGMENTS.}

\author{
         A.S.~BOTVINA}

\address{
 GANIL, 14076 Caen, France
{\em and}\\ Institute for Nuclear Research, 117312 Moscow, Russia.}

\maketitle\abstracts{
The Markov chain statistical multifragmentation model predicts inhomogeneous 
distributions of fragments and their 
isospin in the freeze-out volume caused by an angular momentum and 
external long-range Coulomb field. These effects can take place in peripheral 
nucleus-nucleus collisions at intermediate energies and lead to neutron-rich 
isotopes produced in the midrapidity kinematic region of the reactions.}

Studies of multifragmentation phenomenon in heavy--ion reactions at 
high energies are very promising because of overlapping 
nuclear physics with universal physical processes taking place in finite 
particle systems. In particular, nuclear equations of state and phase 
transitions can be established \cite{hirschegg}. As other complicated 
many-body 
processes this phenomenon can be successfully treated in statistical 
way \cite{PR95,gross97}: 
Fragment production in both peripheral and central collisions  
has clear statistical features \cite{PR95,dagostino99,dagostino96}, 
though a  considerable preequilibrium emission and collective energy (radial 
flow) should be taken into account. 
In finite-size nuclear systems statistical processes can lead to 
unusual effects since the fragment formation is governed 
by both short-range nuclear forces and long-range Coulomb forces. 
For example, a Coulomb interaction of the target and projectile-like sources 
leads to a predominant midrapidity ("neck"-like) emission of intermediate mass 
fragments (IMF, charges $Z$=3--20) \cite{botvina99}. 
In this contribution I show that a statistical 
process can also provide a non-isotropic fragment isospin production in 
peripheral nucleus--nucleus collisions.

The statistical multifragmentation model (SMM) is described in detail in many 
publications \cite{PR95}. 
The model is based upon the assumption of statistical equilibrium at a 
low-density freeze-out stage. 
%Primary fragments are treated as 
%heated nuclear liquid  drops. 
%We consider 
%break-up channels composed of nucleons and excited fragments of different
%masses. 
All possible break-up channels (partitions into fragments) are considered 
with weights defined by the entropies of the channels which depend on 
%taking into 
%account mass, charge, momentum and energy conservations.In the microcanonical 
%treatment the statistical weight of decay channel j is given by 
%$W_{j} \propto exp~S_{j}$, where $S_{j}$ is the entropy of 
%the system in channel $j$ depending on 
excitation energy $E_s^{*}$, 
mass number $A_s$, charge $Z_s$ and other parameters of the source. 
%Light fragments with mass number $A\leq 4$ are considered as stable 
%particles ("nuclear gas") with only translational degrees of freedom; 
%fragments with $A > 4$ are treated as heated nuclear liquid  drops. 
%Different break-up partitions are initialized according to their statistical 
%weights uniformly in the phase space. 
After break-up of the nuclear source 
the fragments propagate independently in their mutual Coulomb fields and
undergo secondary decays. 
%The deexcitation of the hot primary fragments
%proceeds via evaporation, fission or Fermi-break-up\cite{botvina87}. 
The new version of SMM version is based on producing the Markov chain 
of partitions which exactly characterize the whole partition 
ensemble \cite{mc_pre}. 
In a special way individual partitions are generated and selected into the 
chain by applying the Metropolis receipt \cite{mc_pre,mc_smm}. 
%In some aspects it is 
%similar to Metropolis procedure used in Ref. \cite{gross97}. 
Within this 
method primary hot fragments can be placed directly into the freeze-out 
volume to calculate their Coulomb interaction and moment of inertia. 
In this way one can take into account the 
correlations between positions of the primary fragments and their Coulomb 
energy that influences the partition probabilities. 
Angular momentum conservation can be included within this method similar to 
Refs.\cite{gross97,bot_gro95}. 
%In the following the new version 
%\cite{mc_pre,mc_smm} is called the Markov chain SMM. It combines 
%such an advantage of the model \cite{gross97} as an explicit 
%generation of fragments coordinates with accounting experimentally proved 
%properties of the hot fragments suggested in the SMM \cite{PR95}. 
The full analysis of the Markov chain SMM appears somewhere \cite{mc_smm}. 
%here I report about findings which are important in view of broad 
%discussion about isospin influence on fragmentation process \cite{hirschegg}. 

\begin{figure}[tbh]
\begin{minipage}[t]{50mm}
\epsfig{figure=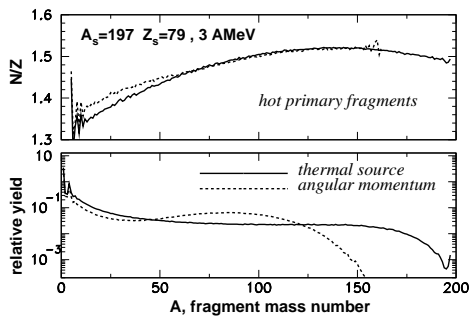,height=4.9cm}
\end{minipage}
%\centerline{\epsfig{figure=bol2000f1_b.eps,height=6.5cm}}
%\includegraphics*[scale=0.8]{bol2000f2.eps}
\hfill
\begin{minipage}[t]{47mm}
\vspace{-43mm}
\caption{The neutron-to-proton ratio N/Z and relative yield of 
hot primary 
fragments produced in the freeze-out after break-up of Au nucleus.
Solid lines: Markov chain SMM calculations for a thermal source with 
excitation 
energy 3 MeV/nucleon, dashed lines: the same source with angular momentum 
150$\hbar$.}
\end{minipage}
\end{figure}

An angular momentum influence on isospin of fragments emitted from a single 
source is instructive. In Fig.~1 I show yields and neutron--to--proton 
(N/Z) ratios of hot primary fragments produced in the freeze-out volume 
(the density is $\rho_s$=$\rho_0$/6, $\rho_0$ is normal nuclear density) 
after break-up of Au source ($A_s$=197, $Z_s$=79) at $E_s^{*}$=3 
MeV/nucleon. It is seen that an angular momentum favors fission-like 
fragment partitions with two large equal-size fragments (see also 
Refs.\cite{gross97,bot_gro95}). That is different from a normal fragmentation 
pattern dominated by partitions with different-size fragments. An angular 
momentum leads to increasing N/Z ratio of IMF also. 
The last effect is important and has a simple 
qualitative explanation: An angular momentum favors emission of IMF with 
larger mass numbers since the system in the freeze-out needs to have a 
large moment of inertia in oder to minimize rotational energy 
and maximize the entropy. From 
another side a Coulomb interaction prevents to emit IMF with large charge 
$Z$. As a result of interplay of these two factors we obtain the increasing 
of the N/Z ratio. 

In peripheral nucleus--nucleus collisions at the projectile energies of 
10--100 MeV/nucleon a break-up of highly excited projectiles-like nuclei is 
fast (the characteristic time is around 100 fm/c) and happens in 
the vicinity of the target-like nuclei. The influence of the Coulomb field of 
the target nucleus on fragmentation of the projectile source 
increases charge asymmetry of produced fragments and leads to non-isotropic 
fragments emission: small fragments are preferably emitted to the side of 
the target \cite{botvina99}. Within the Markov chain SMM one can study how 
this effect influences the isotope composition of fragments.

Calculations were performed for the same Au source as in Fig.~1. 
The source was placed at a fixed distance (20 fm) from another Au source. This 
distance was obtained under assumption that the break-up happens in 
$\sim$100 fm/c after a peripheral collision of 35 $A\cdot$MeV projectile Au 
with target Au. 
%The assumption of a fixed distance between the 
%sources is simplification. 
It is naturally to expect the decays happen at different distances, excitation 
energies and angular momenta. In statistical approach we can take 
into account a distribution of the sources in distances and other 
characteristics by considering an {\it ensemble} of the 
sources. Parameters of this ensemble can be found by global comparison with 
the experiment \cite{PR95,deses98}. However, the present approximation of a 
fixed distance is sufficient for qualitative identification of new 
statistical effects. 

\begin{figure}[tbh]
%\begin{center}
\begin{minipage}[t]{6.8mm}
\epsfig{figure=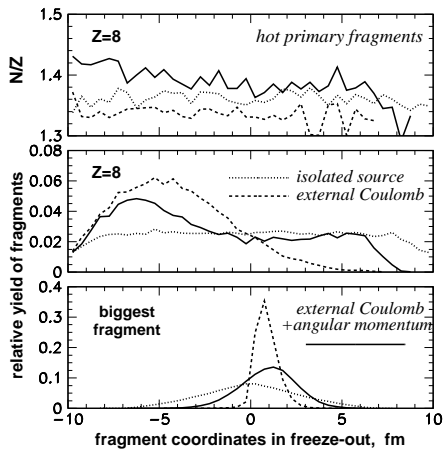,width=6.8cm}
%\end{center}
\end{minipage}
\hfill
\begin{minipage}[t]{50mm}
\vspace{-60mm}
\caption{Freeze-out volume coordinate distributions of 
neutron-to-proton 
ratio N/Z of primary fragments with Z=8 (top panel) and relative yields of 
the primary Z=8 and biggest fragments (middle and bottom panels) produced 
at break-up of Au source at excitation energy 3 MeV/nucleon. 
The second Au nucleus is placed at -20 {\em fm} from the center of the 
freeze-out. 
Dotted lines: the isolated Au source, dashed lines: Coulomb influence 
of the second Au is included, solid lines: angular momentum 150 $\hbar$ is 
included additionally.}
\end{minipage}
\end{figure}

In Fig.~2 I show distributions of yields and N/Z ratio for hot primary 
IMF with $Z$=8 and the biggest fragments in the freeze-out volume along the 
axis connecting the projectile and target sources. It is seen that in case 
of a single isolated source all distributions in the freeze-out are 
symmetric respective to the center mass of the source. In case of the 
target Coulomb influence the IMF are mainly produced closer 
to the target while the biggest fragments are shifted to the opposite 
direction. 
These locations of fragments provide minimum of Coulomb energy in the 
target-projectile system. 
However, the external Coulomb alone influences hardly the fragment isospin 
distribution. In case of angular momentum the N/Z ratio of the IMF increases 
considerably and becomes larger when the IMF are closer to the target. 
The reason is again an interplay of the Coulomb and rotational energy: 
The system needs more heavy IMF to have a large moment of inertia while the 
Coulomb energy of the system depends also on IMF distance from the 
target and this energy is lower when the IMF charge is small.

This asymmetry of the IMF isospin distribution survives after secondary 
deexcitation of hot fragments. The following Coulomb propagation push 
the IMF in the direction of the target providing predominant population 
of the midrapidity kinematic region by neutron-rich fragments. The Coulomb 
repulsion may be not sufficient to accelerate fragments up to high energy, 
however, it can fill with the fragments a considerable part of the 
midrapidity region \cite{botvina99}. Within the statistical 
picture a slight radial flow can supply the IMF with high velocities to 
populate the center of the midrapidity zone.

In conclusion, it was shown that in peripheral nucleus--nucleus collisions 
characteristics of statistically produced fragments depend on Coulomb 
interaction between the target- and projectile-like sources and an angular 
momentum transferred to the sources. In particular, it leads to space 
asymmetry of 
both fragment emission and their isotope composition respective to the 
sources. Previously the 
symmetry violation was considered as a sign of a dynamical "neck" emission. 
However, there is an alternative statistical explanation: the symmetry of the 
phase space is deformed under interaction of the two sources. 
Theoretically such a process gives an example of a new kind 
of statistical phenomenon influenced by an inhomogeneous external long--range 
field \cite{botvina99}. 

%The author thanks GANIL for financial support and warm hospitality. 

\end{document}